\documentclass[shortnote,twocolumn]{jpsj3}
\usepackage{txfonts}
\usepackage[dvipdfmx]{graphicx}
\usepackage{color}
\usepackage{dcolumn}
\usepackage{bm}
\usepackage{amsmath,amssymb}
\usepackage{ulem}
\usepackage{setspace}

\begin{document}
\title{Reduction annealing effects on the crystal structure of $T^{\prime}$-type La$_{1.8}$Eu$_{0.2}$CuO$_{4+\alpha-\delta}$}

\author{Masaki Fujita$^{1}$\thanks{fujita@imr.tohoku.ac.jp}, 
Takanori Taniguchi$^{1}$, Shuki Torii$^2$, Takashi Kamiyama$^{2,3}$,  
Koki Ohashi$^{4}$, Takayuki Kawamata$^{4}$, Tomohisa Takamatsu$^{4}$, Tadashi Adachi$^{5}$, Masatsune Kato$^{4}$, Yoji Koike$^{4}$}
\inst{
$^1$Institute for Materials Research, Tohoku University, Katahira, Sendai 980-8577, Japan\\
$^2$Institute of Materials Structure Science, High Energy Accelerator Research Organization, Tsukuba, Ibaraki 305-0801, Japan\\
$^3$Materials and Life Science Division, J-PARC Center, Tokai, Ibaraki 319-1195, Japan\\
$^4$Department of Applied Physics, Tohoku University, Sendai 980-8579, Japan\\
$^5$Department of Engineering and Applied Sciences, Sophia University, Chiyoda, Tokyo 102-8554, Japan.
} 

\abst{
We performed neutron powder diffraction measurements on the as-sintered (AS) and oxygen-reduced (OR) La$_{1.8}$Eu$_{0.2}$CuO$_{4+\alpha-\delta}$ (LECO). The structural parameters for oxygens in AS and OR samples refined by the Rietveld analysis are almost identical to those of the reference system, Pr$_{2}$CuO$_{4+\alpha-\delta}$. Thus, the two systems are comparable in terms of the structural changes in oxygens due to the annealing, although superconductivity appears only in LECO. 
}

\maketitle

$RE_2$CuO$_4$ ($RE$: trivalent rare earth element) with the Nd$_2$CuO$_4$-type ($T^{\prime}$-type) structure has long been considered as a Mott insulator that is analogous to La$_2$CuO$_4$ with the K$_2$NiF$_4$-type structure. The discovery of superconductivity in the thin film of $T^{\prime}$-type $RE_2$CuO$_4$, thus, has introduced a new aspect, i.e. oxygen coordination around Cu ions, to the study of cuprate superconductors~\cite{Tsukada2005}. 
Superconductivity in the parent $T^{\prime}$-type $RE_2$CuO$_4$ (undoped superconductivity) due to the adequate annealing has been observed in the low temperature (LT) synthesized powder samples~\cite{Asai2011, Takamatsu2012}. It has been discussed that partially existing excess oxygen ions below and above Cu ions (apical oxygen ions) prevent the superconducting (SC) transition in the as-sintered (AS) samples and can be removed by the annealing\cite{Radaelli1994}. The absence of superconductivity in $RE_2$CuO$_4$ synthesized by the conventional high-temperature (HT) method is thought to be due to the difficulty in the complete removal of apical oxygen ions. Therefore, to understand the mechanism of the undoped superconductivity, the structural information of oxygen ions is indispensable. However,  the positional parameters and occupancy of each oxygen site are still unknown for the SC $RE_2$CuO$_4$. 

In this short note, we present the result of neutron powder diffraction on $T^{\prime}$-type La$_{1.8}$Eu$_{0.2}$CuO$_{4+\alpha-\delta}$ (LECO), which exhibits superconductivity below 20 K due to the annealing~\cite{Takamatsu2012, Sunohara2020}. Here, $\alpha$ and $\delta$ are amounts of excess oxygen from the stoichiometric composition and removed oxygen due to the annealing, respectively. We also report the structural information of Pr$_{2}$CuO$_{4+\alpha-\delta}$ (PCO), which remains a non-superconductor even after the annealing, as reference. Comparing these two systems is expected to yield clues for the mechanism of the undoped superconductivity from the structural point of view.  

AS powder of LECO and PCO were prepared by the LT synthesis and solid-state reaction, respectively. LECO were annealed in vacuum at 700$^{\circ}$C for 24 h, whereas PCO  were annealed in Ar-gas flow at 900$^{\circ}$C for 12 h to obtain the oxygen-reduced (OR) sample. The details are described elsewhere~\cite{Takamatsu2012, Sunohara2020, Asano2018}. 
Neutron diffraction measurements with the time-of-flight method were carried out on SuperHRPD at Japan Proton Accelerator Research Complex (J-PARC)~\cite{Torii2011}. We placed the samples in a vanadium cell and collected data at room temperature ($\sim$295 K). Rietveld refinement was performed to evaluate the structural parameters using a Z-Rietveld program~\cite{Oishi2009}. We conducted supplementary experiments on HERMES at Japan Research Reactor No. 3 (JRR-3). 

	\begin{figure}[b]
	\begin{center}
	\includegraphics[width=55mm]{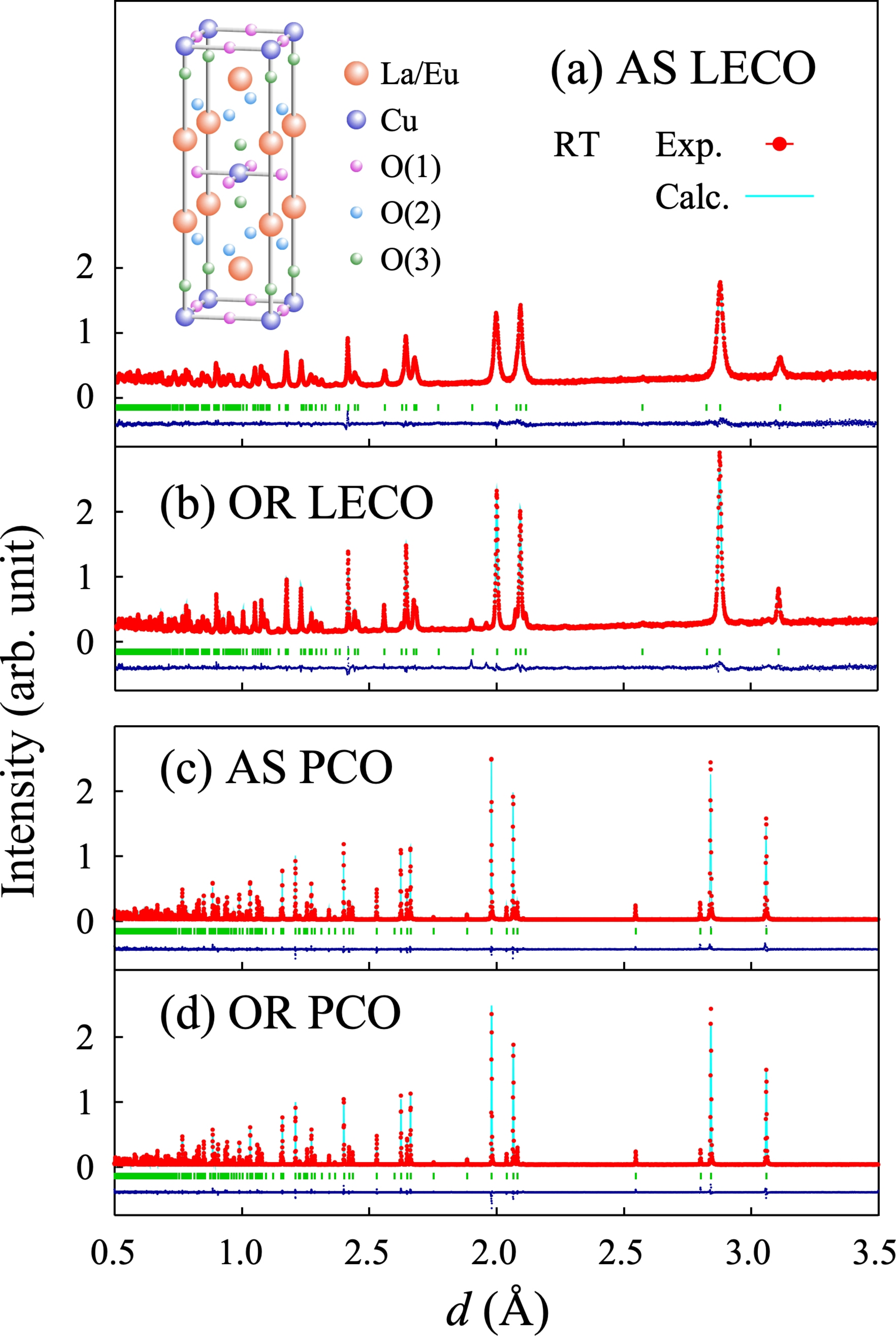}
	\caption{(Color online) Powder diffraction patterns of (a) as-sintered (AS) La$_{1.8}$Eu$_{0.2}$CuO$_{4+\alpha-\delta}$ (LECO), (b) oxygen-reduced (OR) LECO, (c) AS Pr$_{2}$CuO$_{4+\alpha-\delta}$ (PCO) and (d) OR PCO, as a function of lattice spacing $d$. Light blue solid lines (green vertical lines) are calculated patterns (position of reflections) by the Rietveld refinement. Blue marks indicate the difference. The inset depicts a schematic picture of the crystal structure of LECO.}
	\label{profile}
	\end{center}
	\end{figure}

Figures \ref{profile}(a) and (b) show the powder diffraction pattern of AS and OR LECO, respectively, as a function of lattice spacing $d$. No evidence of impurity phase was detected. Thus, we performed Rietveld refinement with a single tetragonal phase characterized by $I$4/$mmm$ symmetry. The peak width was sharper in the OR sample, implying the increase in a crystal grain size due to the annealing. In the analysis, we considered three oxygen sites (see inset) and assumed the followings: (1) The occupancy of La and Eu ions is 1.8 and 0.2, respectively, which is the same as a nominal ratio; (2) the atomic displacement parameters ($B$-factors) for La and Eu ions are the same, and (3) the values of $B$-factors for the three oxygen sites are equal. The neutron absorption effect of Eu ions on the scattering intensity was corrected by considering the shape and density of the sample. We first attempted the fitting with the occupancy of all atoms as a free parameter. Consequently, the occupancy for O(1) and Cu exceeded 2 and 1 respectively for both samples. Then, we fixed them at the full occupancy and refined the parameters again. As can be seen in Figs. \ref{profile}(a) and (b), the diffraction patterns for both samples were satisfactory.

The structural parameters are summarized in Table \ref{para}. The $a$-axis is slightly elongated and the $c$-axis is shrunken due to the annealing~\cite{Sunohara2020}, while the $z$ value for La/Eu and O(3) is not affected by the annealing. 
Notably, finite apical oxygen ions remain in the OR sample, while the amount of apical oxygen decreases by 0.03 due to the annealing. By contrast, a slight defect exists in the $RE_2$O$_2$ layers, and this defect is almost unchanged by the annealing. These results suggest that the variation of oxygen due to the annealing is predominantly confined to the reduction of apical oxygen which is consistent with the neutron diffraction studies of Nd$_2$CuO$_4$ single crystals~\cite{Radaelli1994}. 
From the refinement results, we evaluated $\alpha$ $\sim$ 0.02 and $\delta$ $\sim$ 0.05 for the present LECO~\cite{calc}. 
%

\begin{table}[b]
 \caption{Positional parameters and occupancies per formula unit of La$_{1.8}$Eu$_{0.2}$CuO$_{4+\alpha+\delta}$ and Pr$_{2}$CuO$_{4+\alpha+\delta}$ refined by the Rietveld analysis on neutron diffraction data. The space group is $I$4/$mmm$ (space group No. 139) with the following atomic positions: La/Eu and Pr at 4$e$ (0, 0, $z$), Cu at 2$a$ (0, 0, 0), O(1) at 4$c$ (0, 1/2, 0), O(2) at 4$d$ (0, 1/2, 1/4), and O(3) at 4e (0, 0, $z$). Wyck. and Occ. represent the Wyckoff position and the occupancy of each atom per formula unit. $R_{\rm wp}$ and $R_{\rm p}$ are the reliability factors.}
 \label{table:crystal structure}
 \footnotesize  
 \centering
 \begin{spacing}{0.9}
  \begin{tabular}{clllll}
   \hline
   LECO& & as-sinterd  & & annealed\\
   \hline 
   $a$(\AA) & & 4.0000(1)& & 4.0038(1)&\\
   $c$(\AA) & & 12.4806(1)& & 12.4577(4)&\\
 \hline 
 Atom & Wyck. & $z$ & Occ. & $z$ & Occ.\\
   \hline 
   La/Eu & 4$e$ & 0.352(2) & 1.8/0.2 & 0.353(2) & 1.8/0.2\\
   Cu & 2a &  & 1 &  & 1\\
   O(1) & 4$c$ &  & 2 &  & 2\\
   O(2) & 4$d$ &  & 1.96(4) &  & 1.94(8)\\
   O(3) & 4$e$ & 0.076(3) & 0.06(5) & 0.078(3) & 0.03(1) \\
   \hline 
   $R_{\rm wp}$ ($\%$) & & 5.1 & & 6.2&\\
   $R_{\rm p}$ ($\%$) & & 4.2 & & 4.7&\\
        \hline 
   \hline
     PCO& & as-sinterd  & & annealed\\
   \hline 
   $a$(\AA) & & 3.9603(5)& & 3.962(2)&\\
   $c$(\AA) & & 12.2374(3)& & 12.2412(2)&\\
 \hline 
 Atom & Wyck. & $z$ & Occ. & $z$ & Occ.\\
   \hline 
   Pr & 4$e$ & 0.3515(1) & 1 & 0.3518(1) & 1\\
   Cu & 2$a$ &  & 1 & & 1\\
   O(1) & 4$c$ & & 2 & & 2\\
   O(2) & 4$d$ & & 1.99(4) & & 1.98(4)\\
   O(3) & 4$e$ & 0.062(3) & 0.03(1) & - & 0 \\
   \hline 
   $R_{\rm wp}$ ($\%$) & & 7.7 & & 7.3&\\
   $R_{\rm p}$ ($\%$) & & 6.1& & 5.6&\\
   \hline 
   	\label{para}
  \end{tabular}
  \end{spacing}
\end{table}

Diffraction patterns of AS and OR PCO are shown in Figs. \ref{profile}(c) and (d), respectively. The peak width is much sharper than that in LECO, and the annealing effect on the width is negligible. Thus, the wider width reflecting the smaller grain size may be characteristic of the LT synthesized sample. We fixed O(3) occupancy at zero in the refinement for OR PCO, because the occupancy, which was treated as a free parameter, acquires a negative value. As shown in Table I, the positional parameters and occupancies of AS and OR PCO are almost identical to those of LECO, suggesting that the variations in oxygen due to the annealing between LECO and PCO is comparable. Using the values of occupancy for O(1), O(2), and O(3), we obtained $\alpha$ of $\sim$0.02 and $\delta$ of $\sim$0.04. This value of $\delta$ is consistent with that evaluated from the weight loss of the sample due to the annealing~\cite{Asano2018}. 

The agreement in the refined parameters for oxygen between LECO and PCO is essential in discussing the mechanism of superconductivity in the Ce-free $T^{\prime}$-type $RE_2$CuO$_4$. The present result rejects the possibility that a sufficient number of electrons are doped by removing a large amount of oxygen in LECO compared to PCO. Since the oxygen value is non-stoichiometric in the OR sample, electrons would exist in the sample. Considering that stoichiometric samples acts as Mott-insulators, the electron concentration in the OR samples ($n_{\rm e}$ = 2($\delta$ - $\alpha$)) is $\sim$ 0.06 and 0.04 for SC LECO and non-SC PCO, respectively. A study of the angle-resolved photo-emission spectroscopy of Pr$_{1-x}$LaCe$_x$CuO$_{4+\alpha-\delta}$  (PLCCO) reported that the SC dome is located in the $n_{\rm e}$ range of $\sim$0.07 to $\sim$0.25, which is estimated from the Fermi surface volume~\cite{Song2017}. In the present OR LECO, $n_{\rm e}$ is insufficient to exhibit superconductivity on the phase diagram of PLCCO. Therefore, other factors should be responsible for the difference in the ground state between LECO and PLCCO (PCO) after the annealing. On the other hand, both $a$- and $c$-axis lengths are longer in LECO than in PCO because the average ion radius of RE is large. The compound with a longer $a$-axis has a smaller energy of the charge-transfer gap than that of a compound with a shorter a-axis~\cite{Tshukada2006}. Thus, the conductivity in LECO increases more rapidly owing to the electron doping. Furthermore, according to the present result, the apical oxygen ions are more distant from CuO$_2$ planes in LECO (0.95$\AA$) than in PCO (0.76$\AA$), resulting in a minor effect on the superconductivity due to residual excess oxygen ions in the OR sample. Such differences in the axis lengths and the distance between the apical oxygen and CuO$_2$ planes may cause the distinct electronic state in $T^{\prime}$-type $RE_2$CuO$_4$.

\color{black}
In summary, we performed neutron powder diffraction measurement on LECO and PCO and Rietveld refinement on the diffraction patterns. We confirmed no significant structural difference in oxygen ions between AS LECO and AS PCO and between OR LECO and OR PCO. Therefore, the two systems are comparable in terms of structural changes in oxygen ions due to the annealing.\\

 \footnotesize  
\hspace{0mm} {\bf Acknowledgements: } \hspace{2mm} 
We thank M. Ohkawara for his technical support at HERMES. The experiments at the J-PARC and JRR-3 were performed under a user program (Proposal Nos. 2014A0209 and 2014B0249) and managed by the Institute for Solid State Physics, the University of Tokyo (Proposal No. 10412
), respectively. This work was partially performed under the GIMRT Program of the Institute for Materials Research, Tohoku University (Proposal Nos. 17K0066, 19K0062, 20N0012, 202012-CNKXX0011).


\begin{thebibliography}{9}
\bibitem{calc} The total amount of oxygen ions in the unit formula ($y$) is $\sim$(2 + 1.96 + 0.06) $\approx$ 4.02 for AS LECO and $\sim$(2 + 1.94 + 0.03) $\approx$ 3.97 for OR LECO, giving the value of $\alpha$ = $y$(AS) - 4 $\approx$ 0.02 and $\delta$ $\approx$ 4.02 - 3.97 $\approx$ 0.05.
\end{thebibliography}
\end{document}